\documentclass[
reprint,
amsmath,amssymb,aps,
]{revtex4-1}

\newtheorem{theorem}{Theorem}
\newtheorem{proposition}{Proposition}
\newtheorem{lemma}{Lemma}
\newtheorem{corollary}{Corollary}
\newtheorem{remark}{Remark}

\def\ba{\begin{array}}
\def\ea{\end{array}}
\def\be{\begin{equation}}
\def\ee{\end{equation}}

\def\ds{\displaystyle}

\def\0{{\bf 0}}
\def\1{{\bf 1}}
\def\2{{\bf 2}}
\def\3{{\bf 3}}
\def\4{{\bf 4}}
\def\5{{\bf 5}}
\def\6{{\bf 6}}
\def\7{{\bf 7}}
\def\8{{\bf 8}}
\def\9{{\bf 9}}

\usepackage{graphicx}\usepackage{dcolumn}\usepackage{bm}\usepackage{amsmath}\usepackage{amscd}\usepackage{amsfonts}\usepackage{extarrows}
\usepackage{multirow}  \usepackage[all,cmtip]{xy}
\usepackage{longtable} \usepackage{epstopdf}
\usepackage{tikz}
\usepackage{etoolbox}
\usepackage{enumitem}
\newcommand*{\circled}[1]{\lower1.7ex\hbox{\tikz\draw (1pt, 1pt)%
    circle (0.9em) node {\makebox[2em][c]{\small #1}};}}
\robustify{\circled}
\usepackage{qcircuit}
\usepackage{subfigure}

\makeatletter
\def\hlinewd#1{%
\noalign{\ifnum0=`}\fi\hrule \@height #1 %
\futurelet\reserved@a\@xhline}
\makeatother

\def\bt{\begin{theorem}}
\def\et{\end{theorem}}
\def\bp{\begin{proposition}}
\def\ep{\end{proposition}}
\def\bc{\begin{corollary}}
\def\ec{\end{corollary}}
\def\bo{\begin{proof}}
\def\eo{\end{proof}}
\def\bx{\begin{example}}
\def\ex{\end{example}}
\def\br{\begin{remark}}
\def\er{\end{remark}}
\def\bl{\begin{lemma}}
\def\el{\end{lemma}}

\newcommand{\bea}{\begin{eqnarray}}
\newcommand{\eea}{\end{eqnarray}}
\newcommand{\beas}{\begin{eqnarray*}}
\newcommand{\eeas}{\end{eqnarray*}}
\newcommand{\qed}{$\hfill{\square}$}

\begin{document}

\preprint{APS/123-QED}
\title{Data classification by quantum radial basis function networks}

\author{Changpeng Shao}
\email{changpeng.shao@bristol.ac.uk}
\affiliation{School of Mathematics, University of Bristol,  UK
}
\date{\today}

\begin{abstract}
Radial basis function (RBF) network is
a third layered neural network that is widely used in function approximation and data classification.
Here we propose a quantum model of the RBF network.
Similar to the classical
case,
we still use the radial
basis functions as the activation functions.
Quantum linear algebraic techniques and coherent states can be applied to implement these functions.
Differently, we define the
state of the weight  as
a tensor product of single-qubit states. This gives a simple approach
to implement the quantum RBF network
in the quantum circuits.
Theoretically, we prove that
the training is almost quadratic faster than the classical one.
Numerically,
we demonstrate that the quantum RBF network can solve binary
classification problems as good as the classical RBF network.
While the time used for training is much shorter.
\end{abstract}

\pacs{Valid PACS appear here}
\maketitle

Neural networks are important approaches in machine
learning, which have wide applications in many disciplines \cite{Haykin,lecun2015deep}.
With the developments of quantum  techniques, different kinds
of generalizations of neural networks have been investigated in the quantum computer \cite{beer2020training,killoran2019continuous,wan2017quantum,Farhi-Neven,schuld2020circuit,shao2020quantum,cong2019quantum,verdon2018universal}.
One type of neural network has been extensively studied in the quantum case
is the feed-forward neural network (or multilayer perceptron).
For instance, Beer et al. \cite{beer2020training} generalize this by describing the input
as a density operator,
viewing the unitary and  partial
trace operations as the activation functions.
Killoran et al. \cite{killoran2019continuous} consider it in the architecture of
continuous-variable quantum computer.
In this model, the Gaussian and non-Gaussian gates are used to simulate the key operations of classical neural networks. It can implement nonlinear operations while remaining unitary.
Wan et al. \cite{wan2017quantum} make classical neurons implementable in a quantum computer by adding an ancillary qubit.
For near-term applications,
\cite{Farhi-Neven,schuld2020circuit} build feed-forward quantum neural networks using the idea of variational quantum circuits.
Another direction is to use quantum linear algebraic techniques to accelerate the efficiency of
classical neural networks \cite{kerenidis2019quantum,shao,rebentrost2018quantum}. These works mainly show that
the training procedures of certain
neural networks can be accelerated in a quantum computer.

In this paper, we will propose a quantum model of the radial basis function (RBF) networks.
RBF network was
originated from the performing of exact interpolation of a set of data points \cite{broomhead1988radial,broomhead1988multivariable}.
It is a shallow linear neural network which has a simple mathematical
structure and a relatively cheap training procedure.
The network architecture is also similar to the regularization network. From the viewpoint of approximation theory,
RBF network can approximate any multivariate continuous function on a compact domain with high accuracy.
It has the best-approximation property, and the approximation solution is often optimal.
This makes it a popular
alternative of multilayer perceptrons.
Until now many applications of RBF networks have been found, such as function approximation \cite{park1991universal}, data classification \cite{rifkin2002everything},
time series prediction \cite{yee1999dynamic}, and
the representation of  wave function of quantum-mechanical systems
\cite{teng2018machine}.

The architecture of an RBF network is simple.
It is a type of feed-forward neural network that only has one hidden layer.
The activation functions are known as the
radial basis function. The output is a
linear combination of the training weights
and the radial basis functions of the input
vectors.
RBF network builds on the theorem Cover \cite{cover1965geometrical}, which states that a complex pattern-classification
problem, cast in a high-dimensional space nonlinearly, is more likely to be linearly
separable than in a low-dimensional space, provided that the space is not densely populated.
Because of this, the input vectors are usually
mapped to high dimensional spaces by the radial
basis functions.
This idea is widely used in the kernel method, in which the
radial basis functions play the role of feature maps.

Our quantum generalization of the RBF networks
will mainly focus on their ability in data classification.
To fit them into the quantum circuits, we will first make some
changes to the choices of the weights in the RBF network.
Traditionally, in the hidden layer the weights are introduced independently.
As Cover's theorem suggests, if the hidden
layer has enough neurons, the classification problem becomes linearly separable.
Combining this idea and the requirements of quantum circuits,
we define the weight vector as a tensor product of two-dimensional vectors.
By doing so, the number of parameters used
to determine the weights is exponentially reduced. This also simplifies the implementation
of the RBF network in the quantum circuits.
We can use the same idea
to define an equivalent classical RBF networks. The advantage is that the consuming time used for training is
greatly reduced.
In the quantum computer, the cost can be
further reduced quadratically by the quantum
linear algebraic techniques.

The simplification of the structure also brings a disadvantage,
i.e. it cannot be used to approximate functions, which is another
important application of the traditional RBF network.
Based on our experiments,
when solving binary classification problems,
the performance of
the quantum RBF network and the classical RBF network are close to each other.
But the training time of quantum RBF network is much shorter.
When more training samples are involved, the difference becomes more clear.
Similar to the classical RBF network, the performance of
quantum RBF network increases
when more training samples are used
during training.
Nevertheless, among all the tests, the mean square
errors to approximate functions are
always large. This implies that the quantum RBF network is not a good function approximator.


The paper is organized as follows:
In Section
\ref{Radial basis function network and its generalization}, we briefly review the design of RBF network and introduce a generalization of this neural network.
In Section
\ref{Quantum radial basis function network}, we propose the quantum RBF network and its implementations in quantum circuits.
In Section \ref{Numerical tests}, we perform numerical experiments to verify the ability of quantum RBF networks in solving classification problems.
Finally, in Section \ref{Some related ideas toward the support vector machines} we present
a generalization of the idea to the
support vector machines.

\section{Radial basis function network and its generalization}
\label{Radial basis function network and its generalization}

\subsection{Radial basis function network}
\label{Radial basis function network}

Radial basis function (RBF) network is a type of feed-forward neural network that only has one hidden layer \cite{Haykin,broomhead1988radial}.
It builds on the idea of the kernel method.
The feature maps used in the RBF network are radial basis functions, which are real-valued functions whose
values only depend on the distances between the input vectors and a fixed vector. One typical example is the
Gaussian function $\varphi_{c,\sigma}(x) = \exp(-\|x-c\|^2/2\sigma^2)$, where $c$ is a fixed vector and $\sigma$ is a free parameter. An interesting phenomenon in designing RBF networks is that different choice of the radial basis
functions used in the hidden layer have little influence on the performance \cite{Chen}, so we can just focus
on the Gaussian functions. It is one of the most commonly used feature maps.

\begin{figure}[h]
  \[
  \xymatrix@R=0.2cm @C=1.8cm{
  \circled{$x_1$}    \ar[dr]\ar[r]\ar[dddr]  & \circled{$y_1$} \ar[dr]^{w_1} \\
  \vdots         & \circled{$y_2$} \ar[r]^{w_2}    & \circled{$z$} \\
  \circled{$x_n$}  \ar[ur]\ar[uur]\ar[dr]  & \vdots \ar@{}[ur]|\vdots  \\
                                 & \circled{$y_M$} \ar[uur]_{w_M} \\
  }
  \]
  \caption{The structure of RBF network.}
  \label{RBF network}
\end{figure}
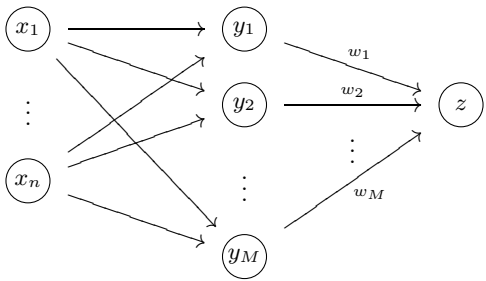

The structure of a RBF network is simple, see Figure \ref{RBF network} for an intuition. The input is an $n$-dimensional real vector $x=(x_1,\ldots,x_n)$,
e.g. the pixel vector of a picture.
In the hidden layer, each
$y_i=\varphi_{c_i,\sigma}(x)$ is determined by a Gaussian function $\varphi_{c_i,\sigma}$. To simplify the design, we use the same $\sigma$ for all the Gaussian functions. Finally, the output is defined as the
linear form $z=\sum_i y_iw_i$ for some weights $w_1,\ldots,w_M$.
More generally, we can add a bias $b$ into the network
and change the output into $b+\sum_i y_iw_i$.

The training of a RBF network contains two steps.
First,  determine the parameters used in the Gaussian functions, i.e. the
centers $c_i$ and the width $\sigma$.
A simple choice of the centers is the training samples. More practically, the centers are found by the $K$-means algorithm.
In this setting, the number of weights can be much smaller than the number of training samples.
In the second step of training, determine the weights by solving a least-square problem.

For instance, consider the binary classification problem.
Suppose we have $M=2^m$ training samples
\be \label{samples}
\{(x^{(t)},r^{(t)})\in \mathbb{R}^n\times \{1,-1\}: t=1,2,\ldots,M\}.
\ee
If the centers are the training samples, then the feature map is defined by
\be\ba{rll} \label{feature map}
f:\mathbb{R}^n &\rightarrow& \mathbb{R}^M \\
x &\mapsto& \left(e^{-\frac{\|x-x^{(1)}\|^2}{2\sigma^2}},\ldots,
e^{-\frac{\|x-x^{(M)}\|^2}{2\sigma^2}} \right).
\ea\ee
Here we choose $\sigma=\max_{r,s} \|x^{(r)}-x^{(s)}\|/\sqrt{2M}$.
The weights $w=(w_1,\ldots,w_M)$ can be obtained by solving the following least-square problem
\be \label{error-eq1}
\min_{w \in \mathbb{R}^M} \quad
\widetilde{L}(w) = \sum_{t=1}^M
\left(f(x^{(t)})\cdot w - r^{(t)} \right)^2.
\ee
The training procedure to solve the above least-square problem is time-consuming for a classical computer.
The recursive least-square method
is often used to reduce the complexity to calculate the matrix inversion.
In a quantum computer, quantum linear solvers can be applied to accelerate
the solving of (\ref{error-eq1}).
This  has been recently investigated in \cite{shao}.
The above is just a brief introduction
about the RBF network, for more
we refer to  \cite{Haykin,orr1996introduction}.

\subsection{A  generalization of RBF network}
\label{A simple generalization of RBF network}

Now we generalize the idea
of RBF network from the choices of the
weights. The goal is to use this
generalization to define a
quantum RBF network.

The basic structure is still the same as Figure \ref{RBF network}. We only change the weights used in the hidden layer into
a tensor product form
\be\label{weight1}
w(\vec{\theta}) = \bigotimes_{j=1}^m
\left( {\begin{array}{r} \vspace{.2cm}
   \cos\theta_j \\
   \sin\theta_j \\
  \end{array} } \right),
\ee
where $m=\lceil \log M \rceil$, and $M$ is the number of training samples.
More precisely, let $t=t_1+t_22+\cdots+t_m2^{m-1}$ be the binary expanding of the integer $t\in\{0,1,2,\ldots,M-1\}$, then
it is easy to verify from equation (\ref{weight1}) that the $t$-th weight equals
\be \label{weight-entry}
w_t = \prod_{j=1}^m \cos(\theta_j-\frac{t_j\pi }{2}).
\ee
Even though the dimension of $w(\vec{\theta})$ is $M$, there are only $\log M$ parameters we need to
determine.

As for the training, similar to
(\ref{error-eq1})
we need to minimize
a loss function
\be \label{error1}
L(\vec{\theta}) = \frac{1}{2M} \sum_{t=1}^M \Big(  f(x^{(t)})  \cdot w(\vec{\theta}) - r^{(t)}\Big)^2,
\ee
where $f$ is defined by equation (\ref{feature map}).
Since only $\log M$ parameters are
need to be determined, the complexity
of minimizing $L$ may not as high as minimizing $\widetilde{L}$.
Note that $L$ does not define a least-square
problem, so we should use general optimization methods like gradient descent or Newton's method to minimize it.
About this, we have the following estimation
about the cost of each iteration  step.
We remark that the following result does not consider other potential problems in the
gradient descent (e.g. the gradient is zero) and in Newton's method (e.g. the Hessian matrix is singular so that regularization is required).
And we will not consider these problems
here as our main goal is to design a quantum RBF network.

\bt
\label{thm}
To minimize $L(\vec{\theta})$, if we use the gradient descent method, then the cost of each step of iteration is $O(nM^2)$.
If we use the Newton's method, then the cost of each step of iteration is $O(M^2(n+\log^2M))$.
\et
{\em Proof.}
First, assume that the gradient descent method is used.
Set the standard basis of $\mathbb{R}^m$ as $\{e_1,\ldots,e_m\}$; i.e., $e_j$ is the $\{0,1\}$ vector such that only the $j$-th entry equals 1.
Then the gradients of $L$ satisfies
\bea
\nabla L &=& \frac{1}{M} \sum_{j=1}^m \Bigg(\sum_{t=1}^M \Big(  w(\vec{\theta}) \cdot f(x^{(t)})  - r^{(t)}\Big) \nonumber \\
&& \hspace{2.1cm} \times \, \frac{\partial w(\vec{\theta})}{\partial \theta_j} \cdot f(x^{(t)}) \Bigg) e_j,  \label{gradient1}
\eea
By equation (\ref{weight-entry}), the $t$-th entry of the vector ${\partial w(\vec{\theta})}/{\partial \theta_j}$ equals
\be \label{weight-entry1}
\prod_{i=1}^m
\cos(\theta_i-\frac{(t_i-\delta^i_j)\pi }{2}),
\ee
where $\delta^i_j$ is the Kronecker symbol.

We can see that the complexity to compute all $f(x^{(t)})$ is $O(nM^2)$.
By equations (\ref{weight-entry}) and (\ref{weight-entry1}),
the complexity to compute
all the entries of $w(\vec{\theta}),$
${\partial w(\vec{\theta})}/{\partial \theta_1},\ldots,$ $
{\partial w(\vec{\theta})}/{\partial \theta_m}$ is $O(m^2M)$. They are only need to be computed once.
After we obtain them, the cost to compute $\sum_{j=1}^m {\partial w(\vec{\theta})}/{\partial \theta_j}$ is $O(mM)$.
The computation of $mM$ inner products in equation (\ref{gradient1}) costs $O(mnM)$.
Thus, the total cost to compute $\nabla L$ is $O(nM^2+m^2M+mnM)=O(nM^2)$.

More advanced optimization methods, such as the Newton's method,
are rarely used in machine learning due
to the high cost to compute the inverse of the Hessian matrix. However, the Hessian matrix
$\nabla^2 L$ of $L$ is $m$-by-$m$.
So the cost of  Newton's method to minimize $L$
is not as high as usual.
More precisely, the $(j,k)$-th entry of $\nabla^2 L$ equals
\beas
&& \frac{1}{M}  \sum_{t=1}^M \Big(  w(\vec{\theta}) \cdot f(x^{(t)})  - r^{(t)}\Big)
\frac{\partial^2 w(\vec{\theta})}{\partial \theta_j\partial \theta_k} \cdot f(x^{(t)}) \\
&& + \, \frac{1}{M}  \sum_{t=1}^M
\left(\frac{\partial w(\vec{\theta})}{\partial \theta_j} \cdot f(x^{(t)}) \right)
\left( \frac{\partial w(\vec{\theta})}{\partial \theta_k} \cdot f(x^{(t)}) \right).
\eeas
Similar to the analysis of computing $\nabla L$,
the cost to calculate the Hessian matrix of $L$ is $O(m^2M^2)$.
On the other hand, the complexity to compute the inverse of $\nabla^2 L$ is $O(m^3)$. Thus the cost of each
iteration of the Newton's method to minimize $L$ is $O(nM^2+m^2M^2+m^3)$.
\qed
\vspace{.2cm}

The above result maybe not surprising as we only
introduce $m=\log M$ parameters. This will
reduce the cost of training but
may weaken the
power of the RBF network in solving practical problems.
In Section \ref{Numerical tests},
we will numerically investigate this.

\section{Quantum radial basis function network}
\label{Quantum radial basis function network}

Based on the generalization of the
RBF network, we now give
a definition of the quantum RBF network.
Although RBF networks can be used to solve many
problems, in the quantum
case we only focus on its ability in data classification. For simplicity, we only give an
explicit description of the quantum RBF network
for solving binary classification problems.

To build a quantum RBF network, there are two
problems we need to solve: First, how to encode the
data into the quantum circuits?
A commonly used strategy is to encode the data
into the amplitudes of a quantum state.
Here we use the Gaussian kernel to map the samples
into a high dimensional space,
then prepare the quantum states of those
high dimensional vectors.
More precisely, for any vector $x\in\mathbb{R}^n$,
we define
\be \label{feature unitary}
|f(x)\rangle := U_x|0\rangle^{\otimes m} \propto \sum_{t=1}^M
\exp\left(- \frac{\|x-x^{(t)}\|^2} {2\sigma^2}\right) \, |t\rangle,
\ee
where $\sigma=\max_{r,s} \|x^{(r)}-x^{(s)}\|/\sqrt{2M}$.
This step is similar to the feature map (\ref{feature map}).

Another problem is how to introduce the training weights.
Based on the idea of the kernel method (or Cover’s theorem), if we map the data
into a high dimensional space, the complex nonlinear
pattern classification problem often becomes linearly
separable. So we can
use linear form to define the weights.
In order to implement it efficiently
in a quantum computer, we use
tensor product to introduce the
linear structure
on the weights.
Thus we define the parameterized unitary as
\be  \label{weight unitary}
U(\vec{\theta}) = \bigotimes_{j=1}^m \exp(iY\theta_j)  = \bigotimes_{j=1}^m
\left( {\begin{array}{rr} \vspace{.2cm}
   \cos\theta_j  & \sin\theta_j \\
   -\sin\theta_j & \cos\theta_j \\
  \end{array} } \right),
\ee
where $Y$ is the Pauli-$Y$ matrix.
As a result, the state $|f(x)\rangle$ is transformed into
\be
|z\rangle := U(\vec{\theta}) |f(x)\rangle = \langle w(\vec{\theta}) | f(x)\rangle \, |0\rangle^{\otimes m} + |0^\bot\rangle^{\otimes m},
\ee
where $|0^\bot\rangle^{\otimes m}$ refers to a state that is orthogonal to the first term, and
\be \label{weight}
|w(\vec{\theta}) \rangle = U^\dag(\vec{\theta}) |0\rangle^{\otimes m} =
\bigotimes_{j=1}^m \Big(\cos\theta_j|0\rangle + \sin\theta_j |1\rangle \Big).
\ee
Finally, we define the output of the quantum RBF network
as the amplitude $\langle w(\vec{\theta}) | f(x)\rangle$
of $|0\rangle^{\otimes m}$ of $|z\rangle$. It can be estimated by Hadamard test in a quantum computer.

\begin{figure}[ht]
\centering
\[
\Qcircuit @C=1em @R=1.3em {
\lstick{|0\rangle}   & \multigate{2}{U_x} & \gate{\exp(iY\theta_1)} & \meter   \\
\lstick{\vdots}   &  & \lstick{\vdots} & \lstick{\vdots}   \\
\lstick{|0\rangle}   & \ghost{U_x} & \gate{\exp(iY\theta_m)} & \meter  \\
} \]
\caption{Quantum circuit of RBF network.}
\label{qnn}
\end{figure}
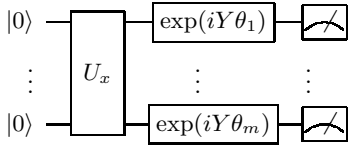

Figure \ref{qnn} describes the quantum circuit of the quantum RBF network. In the following subsections, we will explain more about how to compute the output by Hadamard test,
how to implement $U_x$ and how to train the network.

The structure of the quantum RBF network
is the same as the generalized RBF network
we defined in Section
\ref{A simple generalization of RBF network}.
Because of this, their performance should be close
to each other. However, due to the
quantum linear algebraic techniques, the
calculations in the quantum RBF network
can be done much faster. We will show that this
speedup is almost quadratic.

\subsection{How to compute the output by measurements?}
\label{How to get the output by measurements}

Generally, for any two real unit vectors $x,y$, Hadamard test can only  extract the information $|\langle x|y\rangle|^2$.
A simple technique to overcome this problem is
considering $(1,x)$ and $(1,y)$.
As for the above-defined quantum RBF network, we can use this idea to extract $\langle w(\vec{\theta}) | f(x)\rangle$.

Denote the square of the norm of the vector $f(x)$ as
\be \label{normalization factor}
R=\sum_{t=1}^M \exp(- \|x-x^{(t)}\|^2 / \sigma^2).
\ee
Let
\be
|\widetilde{f}(x)\rangle = \frac{1}{\sqrt{R+1}} \Big(\sqrt{R} \, |0\rangle |f(x)\rangle + |1\rangle |0\rangle^{\otimes m}\Big).
\ee
This state is obtained by introducing one extra sample $x$ into (\ref{samples})
as $\exp(- \|x-x\|^2 / 2\sigma^2)=1$.
To prepare $|\widetilde{f}(x)\rangle$, the label of $x$ is not important.
Consequently, if we know how to
prepare the state
$|f(x)\rangle$, then we can similarly prepare $|\widetilde{f}(x)\rangle$.
This problem is studied in the next subsection.

Now suppose we already have $|\widetilde{f}(x)\rangle$. By viewing the first qubit $|0\rangle$ as the control qubit, we can apply $U(\vec{\theta})$ to $|f(x)\rangle$ to create
\beas
&& \frac{1}{\sqrt{R+1}} \Big(\sqrt{R} \, |0\rangle U(\vec{\theta})|f(x)\rangle + |1\rangle |0\rangle^{\otimes m}\Big) \\
&=& \frac{1}{\sqrt{R+1}} \Big(\sqrt{R}\langle w(\vec{\theta}) | f(x)\rangle |0\rangle|0\rangle^{\otimes m} + |1\rangle|0\rangle^{\otimes m} \\
&& \hspace{1.5cm} +\,{\rm orthogonal~terms}\Big).
\eeas
Apply Hadamard gate to the first qubit, then we obtain
\beas
&& \frac{1}{\sqrt{R+1}} \Bigg(\frac{1}{\sqrt{2}}
\Big(\sqrt{R}\langle w(\vec{\theta}) | f(x)\rangle - 1 \Big) |0\rangle^{\otimes (m+1)}   \\
&& \hspace{1.15cm} \hfill +\, \frac{1}{\sqrt{2}}
\Big(\sqrt{R}\langle w(\vec{\theta}) | f(x)\rangle + 1 \Big) |1\rangle|0\rangle^{\otimes m} \\
&& \hspace{1.15cm} +\,{\rm orthogonal~terms}\Bigg).
\eeas
Apply amplitude estimation to estimate the amplitudes of
$|0\rangle^{\otimes (m+1)}$ and $|1\rangle|0\rangle^{\otimes m}$,
we obtain approximations of
\bea
&& \frac{1}{2(R+1)} \Big(\sqrt{R}\langle w(\vec{\theta}) | f(x)\rangle - 1 \Big)^2, \\
&& \frac{1}{2(R+1)} \Big(\sqrt{R}\langle w(\vec{\theta}) | f(x)\rangle + 1 \Big)^2.
\eea
The difference of these two approximations gives an approximation of $\langle w(\vec{\theta}) | f(x)\rangle$.

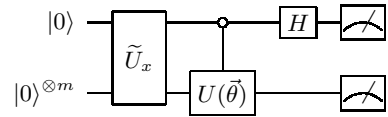
\begin{figure}[ht]
\centering
\[
\Qcircuit @C=1em @R=1.3em {
\lstick{|0\rangle}   & \multigate{1}{\widetilde{U}_x} & \ctrlo{1} & \gate{H} & \meter\\
\lstick{|0\rangle^{\otimes m}}   & \ghost{\widetilde{U}_x} & \gate{U(\vec{\theta})} & \qw& \meter   \\
} \]
\caption{Quantum circuit to compute the output.}
\label{qnn1}
\end{figure}

The above procedure is described in Figure \ref{qnn1}. In the circuit,
$H$ is the Hadamard gate,
$\widetilde{U}_x$ is used to
prepare the state $|\widetilde{f}(x)\rangle$.

\subsection{How to implement the unitary $U_x$?}
\label{The implementation of the unitary}

The input problem is a bottleneck of many quantum machine learning algorithms. In the
design of the quantum RBF network, this is also a constriction.
In this subsection, we will use the idea of coherent states and quantum linear algebraic techniques to implement the unitary $U_x$ in a quantum computer. In the following,
we will consider the preparation of the states $|f(x^{(t)})\rangle$ of
the training samples.
As for the preparation of $|f(x)\rangle$,
we can view $x$ as a new sample and prepare
the state in the same way.

Coherent states play a crucial role in quantum optics and mathematical physics \cite{sanders2012review,fujii2001introduction}. They are defined in the Fock states $\{|0\rangle, |1\rangle, \ldots\}$, which is a basis of the infinitely dimensional Hilbert space $\mathcal{H}$. The number $n$ of $|n\rangle$ represents the number of
phonons.
Let $\mathfrak{a}, \mathfrak{a}^\dag$ respectively be the annihilation, creation operator of the harmonic oscillator, then
\[
\mathfrak{a}|n\rangle = \sqrt{n}|n-1\rangle,
\,\,\,
\mathfrak{a}^\dag|n\rangle = \sqrt{n+1}|n+1\rangle.
\]
For any $n\geq 1$, it is easy to see that
\be \label{operator}
|n\rangle = \frac{(\mathfrak{a}^\dag)^n}{\sqrt{n!}} |0\rangle.
\ee
Let $r\in \mathbb{R}$ be a real number, its coherent state is defined by
\be
|\psi_r\rangle = e^{-r^2/2\sigma^2} \sum_{k=0}^\infty
\frac{(r/\sigma)^k}{\sqrt{k!}} \, |k\rangle.
\ee
It is a unit eigenvector of $\mathfrak{a}$ corresponding to the eigenvalue $r/\sigma$, that is $\mathfrak{a} |\psi_r\rangle = (r/\sigma) |\psi_r\rangle$. From (\ref{operator}),
we also have
$|\psi_r\rangle = e^{-r^2/2\sigma^2}
e^{r\mathfrak{a}^\dag/\sigma}|0\rangle
=e^{r(\mathfrak{a}^\dag-\mathfrak{a})/\sigma}|0\rangle $, thus $|\psi_r\rangle$ is obtained by
a unitary operator of dimension infinity.

As for the preparation of $|\psi_r\rangle$ in a
finite quantum circuit, we can consider its
Taylor approximation:
\be
|\widetilde{\psi}_r\rangle  \propto \sum_{k=0}^{N-1}
\frac{(r/\sigma)^k}{\sqrt{k!}} \, |k\rangle.
\ee
Before normalization, the error
is bounded by
\be \label{error of coherent state}
\sum_{k=N}^{\infty}
\frac{(r/\sigma)^{2k}}{k!}
\leq \frac{(r/\sigma)^{2N}}{N!} e^{(r/\sigma)^2}.
\ee
By Stirling's approximation, $N! \approx \sqrt{2\pi N}(N/e)^N$. Set the upper bound of the error (\ref{error of coherent state}) as $\epsilon$, then
\[
\log \frac{1}{\epsilon} - \log\sqrt{2\pi} + \frac{r^2}{\sigma^2}
\leq (N+\frac{1}{2}) \log N  -
N (1 + 2\log\frac{r}{\sigma}).
\]
So we can choose $N = O(r^2/\sigma^2 + \log 1/\epsilon)$.
With this choice,
$\||\psi_r\rangle-|\widetilde{\psi}_r\rangle\|^2\leq e^{r^2/\sigma^2}\epsilon$.
As for the quantum state $|\widetilde{\psi}_r\rangle$, it can be prepared in time $O(\log 1/\epsilon)$ by a standard technique \cite{shende2006synthesis}.
Similar to the Taylor expanding, another method is to truncate the
skew-Hermitian matrix $\mathfrak{a}^\dag-\mathfrak{a}$ to dimension $O(\log 1/\epsilon)$.
This is a 2-sparse matrix, the Hamiltonian simulation is efficient \cite{berry2007efficient}.
Therefore, there is an efficient quantum circuit to prepare the coherent state $|\psi_r\rangle$ up to precision $\epsilon$ in time $O(\log 1/\epsilon)$.

The coherent state for a real vector $x=(x_1, \ldots, x_n)$ is defined by
\be
|\psi_x\rangle =  |\psi_{x_1}\rangle \otimes \cdots \otimes |\psi_{x_n}\rangle.
\ee
It is not hard to check that for any two real vectors $x,y$,
we have $\langle \psi_x|\psi_y\rangle = \exp(-\|x-y\|^2/2\sigma^2)$.

Now consider the following superposition of coherent states of the training samples:
\be
|\Psi\rangle = \frac{1}{\sqrt{M}} \sum_{t=1}^M |t\rangle |\psi_{x^{(t)}}\rangle.
\ee
This state is obtained by first preparing all the coherent states
$|\psi_{x^{(t)}}\rangle$, then
generate their superposition by Hadamard transform.
Since the error in preparing each state $|\psi_{x^{(t)}}\rangle$ is bounded by $n\epsilon$, to make sure the error
of preparing $|\Psi\rangle$ is bounded by $\delta$, we choose $\epsilon=\delta/n$. Consequently,
the total cost to prepare $|\Psi\rangle$ is $O(Mn\log 1/\epsilon)
=O(Mn\log n/\delta)$.

Taking the partial trace on the second register of $|\Psi\rangle \langle \Psi|$ gives rise to the density operator of the kernel matrix
\be\ba{lll} \vspace{.2cm}
\label{density operator}
\rho &:=& {\rm Tr}_2 |\Psi\rangle \langle \Psi| \\
&=&  \ds \frac{1}{M} \sum_{s,t=1}^M
\exp(-\|x^{(s)}-x^{(t)}\|^2/2\sigma^2) \,
|s\rangle\langle t| .
\ea\ee
As a result,
\be\ba{lll}
\label{matrix-vector multiplication}
\vspace{.2cm}
\rho |t\rangle &=&\ds \frac{1}{M} \sum_{s=1}^M
\exp(-\|x^{(s)}-x^{(t)}\|^2/2\sigma^2) \,
|s\rangle \\
&=&\ds \frac{\sqrt{R_t}}{M}|f(x^{(t)})\rangle,
\ea\ee
where $R_t
=\sum_{s=1}^M \exp(-\|x^{(s)}-x^{(t)}\|^2/\sigma^2)$
is the square of the norm of the vector $f(x^{(t)})$.
The matrix-vector multiplication (\ref{matrix-vector multiplication}) can be
accomplished by the basic quantum linear algebraic techniques, such as the block-encoding \cite{gilyen2019quantum}.
With the above preliminaries,
by Lemma 45 of \cite{gilyen2019quantum}, we can construct an $(1, n+\log M,0)$-block-encoding of $\rho$ in cost $O(Mn\log n/\delta)$. As a result,
we can create
\be \label{input state}
\rho |t\rangle |0\rangle + |0\rangle^\bot
\ee
at the same time.
Note that if we first compute all $\exp(-\|x^{(s)}-x^{(t)}\|^2/2\sigma^2)$, then prepare the
quantum states of the samples, the cost is at least $O(M^2)$,
which is the same as the complexity of the classical
input reading problem.
So the above quantum procedure shows that quantum
RBF can have a quadratic speedup on $M$
in the input problem.

Certainly, we can perform measurements on the state (\ref{input state}) to get the state $|f(x^{(t)})\rangle$.
We can also choose (\ref{input state}) as the result of $U_x$ since we need to do the
amplitude estimation at the end of the quantum circuit in Figure \ref{qnn}.

\subsection{How to train the quantum RBF network?}
\label{How to train the variational quantum classifier}

Following the idea of variational quantum eigensolver \cite{Peruzzo}, the training  of the
quantum RBF network is accomplished by using the quantum-classical hybrid method to minimize the loss function
\be \label{error}
L(\vec{\theta}) = \frac{1}{2M} \sum_{t=1}^M \Big(\langle w(\vec{\theta}) | f(x^{(t)})\rangle - r^{(t)}\Big)^2.
\ee

As discussed in Theorem \ref{thm},
the costs of gradient descent and Newton's
method to minimize $L(\vec{\theta})$
only differ by a logarithmic term at each
step of iteration. In the quantum computer,
this result should also correct.
By the quantum linear algebraic techniques,
in a quantum computer the cost of each iteration of these two methods can be reduced quadratically.
In the following, we only present an
analysis of the gradient descent method
to minimize  $L(\vec{\theta})$.

When considering the training, based on  the algorithm in the above subsection to prepare the quantum states, we can change the quantum circuit shown in Figure \ref{qnn}
into Figure \ref{qnn2} below.
Note that in Figure \ref{qnn2},
$U_{\rm in}$ does not depend on $t$, the index of the $t$-th sample.

\begin{figure}[ht]
\centering
\[
\Qcircuit @C=1em @R=1.0em {
&\lstick{|t\rangle} & \multigate{1}{\mathcal{O}_{\rm id}} & \qw & \qw & \qw \\
&\lstick{|0\rangle^{\otimes m}} & \ghost{\mathcal{O}_{\rm id}} & \gate{\rho} & \gate{U(\vec{\theta})}  &  \meter
 \gategroup{1}{3}{2}{4}{0.7em}{--} \\
&&& \lstick{U_{\rm in}}
} \]
\caption{Quantum circuit of the quantum RBF network, where
$\rho$ is the density operator (\ref{density operator}) and
$\mathcal{O}_{\rm id}|t,0\rangle=|t,t\rangle$.}
\label{qnn2}
\end{figure}
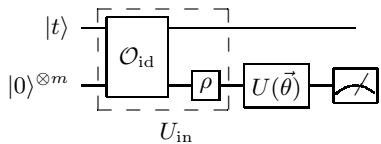

In the gradient descent, we update $\vec{\theta}$ via
\be \label{update}
\theta_j = \theta_j - \eta \sum_{t=1}^M
\Big(\langle w(\vec{\theta}) | f(x^{(t)})\rangle - r^{(t)}\Big)
\langle \frac{\partial  w(\vec{\theta})}{\partial\theta_j} |
f(x^{(t)})\rangle,
\ee
where $\eta$ is the learning step, $j=1,2,\ldots,m.$
By equations (\ref{weight unitary}) and (\ref{weight}),
\be\ba{lll} \label{derivative}
\vspace{.2cm}
\ds \langle \frac{\partial  w(\vec{\theta})}{\partial\theta_j} |
&=&  \ds \langle 0|^{\otimes m}\bigotimes_{l=1}^{j-1} \exp(iY\theta_l) \\
&& \hspace{-.9cm} \ds \otimes\,
\exp(iY(\theta_j+\frac{\pi}{2}))
\otimes
\bigotimes_{l=j+1}^{m}  \exp(iY\theta_j) .
\ea\ee
Consequently, to estimate $\langle \frac{\partial  w(\vec{\theta})}{\partial\theta_j} |
f(x^{(t)})\rangle$, we only need to change the parameter
$\theta_j$ into $\theta_j+\frac{\pi}{2}$ in Figure \ref{qnn2},
then use the amplitude estimation to estimate the amplitude of $|0\rangle^{\otimes m}$.

Now we consider the calculation of
the summation in
equation (\ref{update}).
Denote the unitary in (\ref{derivative}) as $U_j(\vec{\theta})$.
For the first term of the summation,
by (\ref{matrix-vector multiplication}),
it equals
\be
\label{first quantity}
\langle 0|^{\otimes m} U(\vec{\theta})\rho \left(\sum_{t=1}^M \frac{M^2}{R_t}   |t\rangle \langle t| \right)
\rho U_j(\vec{\theta}) |0\rangle^{\otimes m}.
\ee
Notice that
$U(\vec{\theta}),U_j(\vec{\theta})$
are tensor products of single-qubit gates,
$\rho$ can be implemented by
quantum linear algebraic techniques (see (\ref{input state})), and $\sum_{t=1}^M ({M^2}/{R_t}) |t\rangle \langle t|$ is
a diagonal matrix, the quantity (\ref{first quantity})
can be estimated to certain precision
in time linear on $M$.

The second term of the summation of (\ref{update}) equals
\be
\label{second quantity}
\langle 0|^{\otimes m} U_j(\vec{\theta})\rho
\sum_{t=1}^M \frac{Mr^{(t)}}{\sqrt{R_t}}
 | t\rangle.
\ee
The quantum state $\sum_{t=1}^M ({Mr^{(t)}}/{\sqrt{R_t}})
| t\rangle$ can be prepared by the
standard techniques in time $O(M)$.
So the quantity (\ref{second quantity}) can also be computed in
time linear at $M$.
Compared to the result of Theorem \ref{thm},
the training of quantum RBF network is almost quadratic faster than the classical RBF network.

\section{Numerical tests}
\label{Numerical tests}

To test the performance of the quantum RBF network, we use it to solve binary
classification problems in dimension two.
However, we only do the experiments for the generalized RBF
network defined in Section \ref{A simple generalization of RBF network}
in the classical computer since we do not have a
large-scale quantum computer now. Due to the closeness, the testing
results should be true for
a fault-tolerant quantum computer.

\begin{figure}[ht]
\centering
\subfigure[]{
\includegraphics[height=1.7cm,width=2cm]{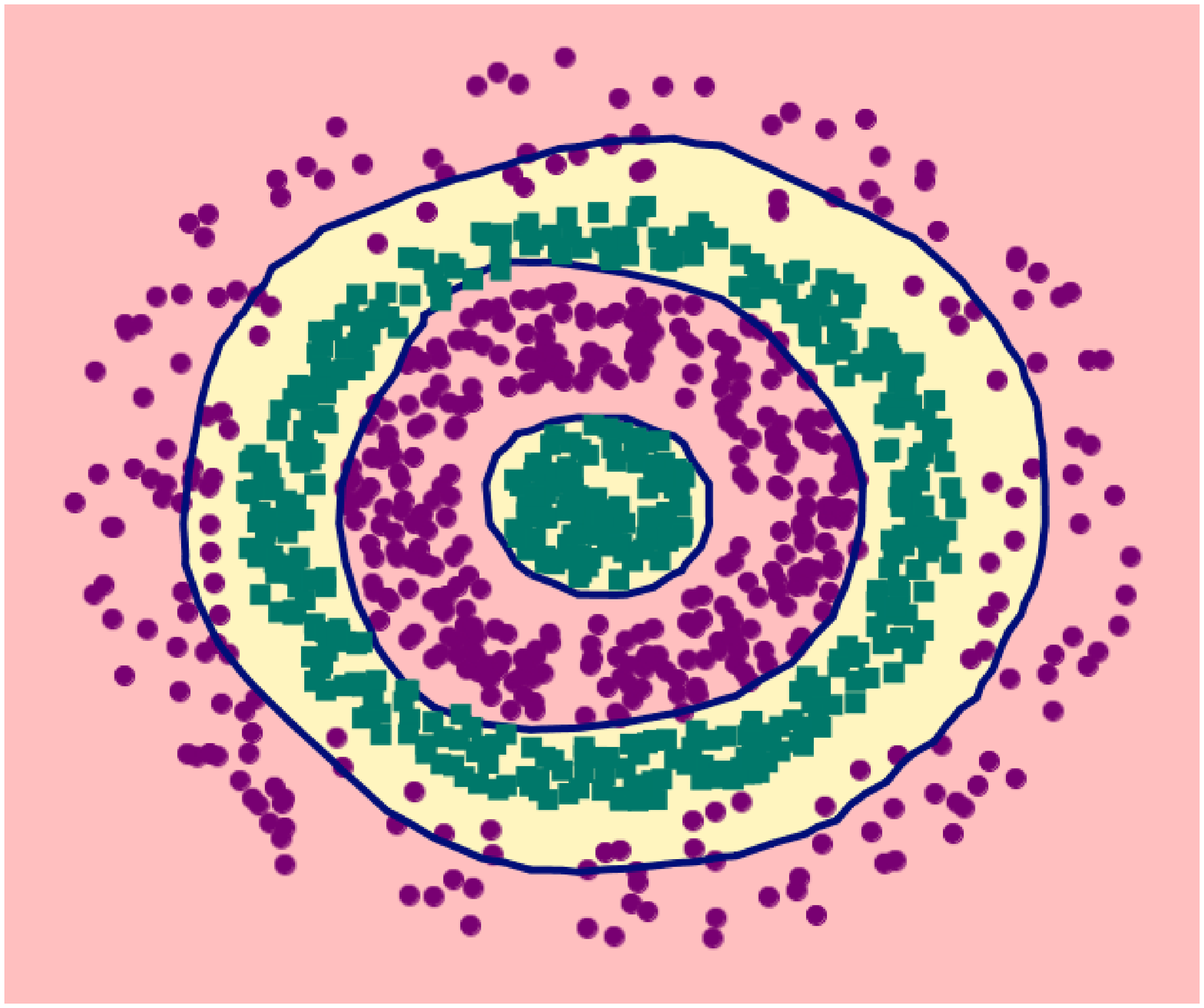}
}\hspace{-.21cm}
\subfigure[]{
\includegraphics[height=1.7cm,width=2cm]{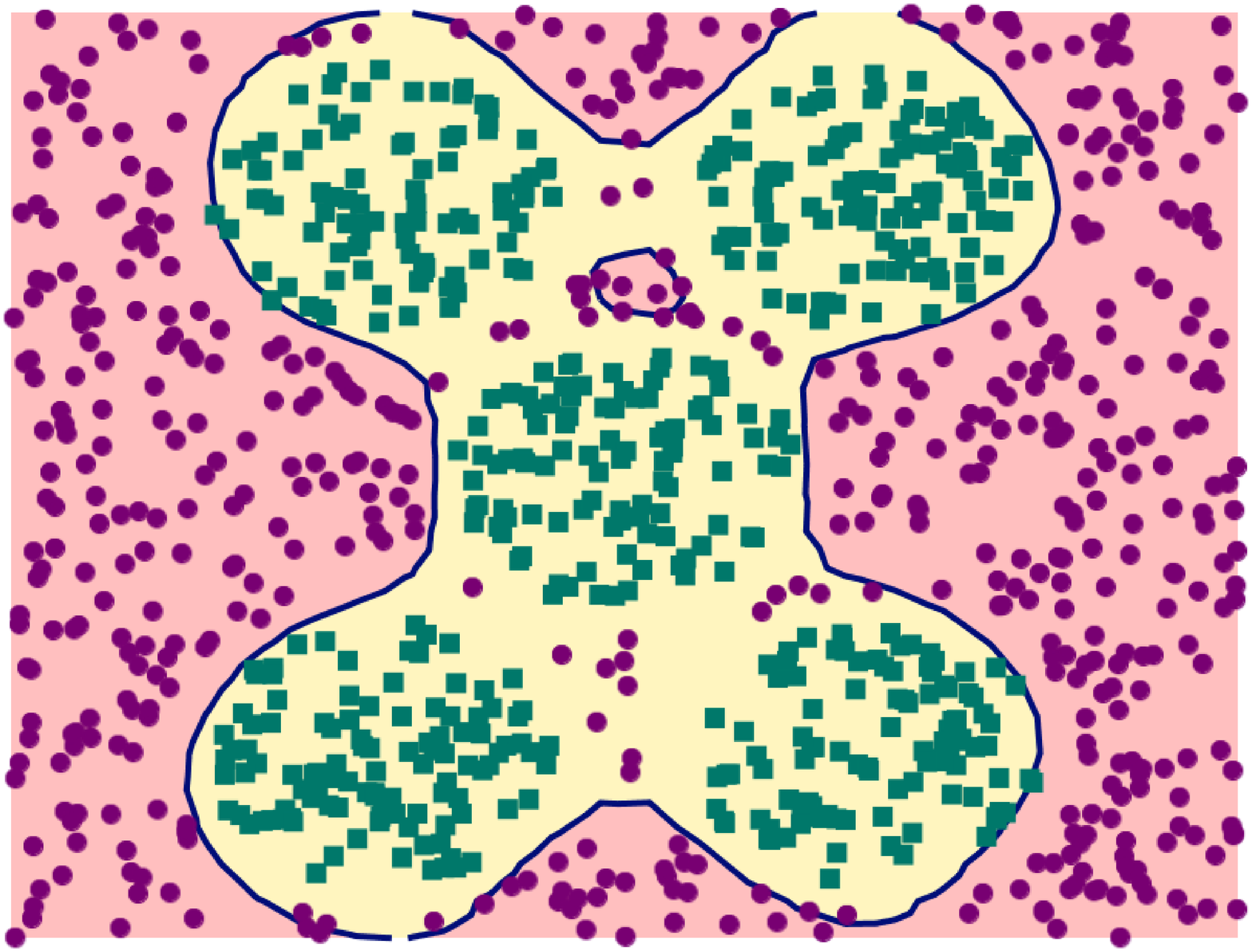}
}\hspace{-.21cm}
\subfigure[]{
\includegraphics[height=1.7cm,width=2cm]{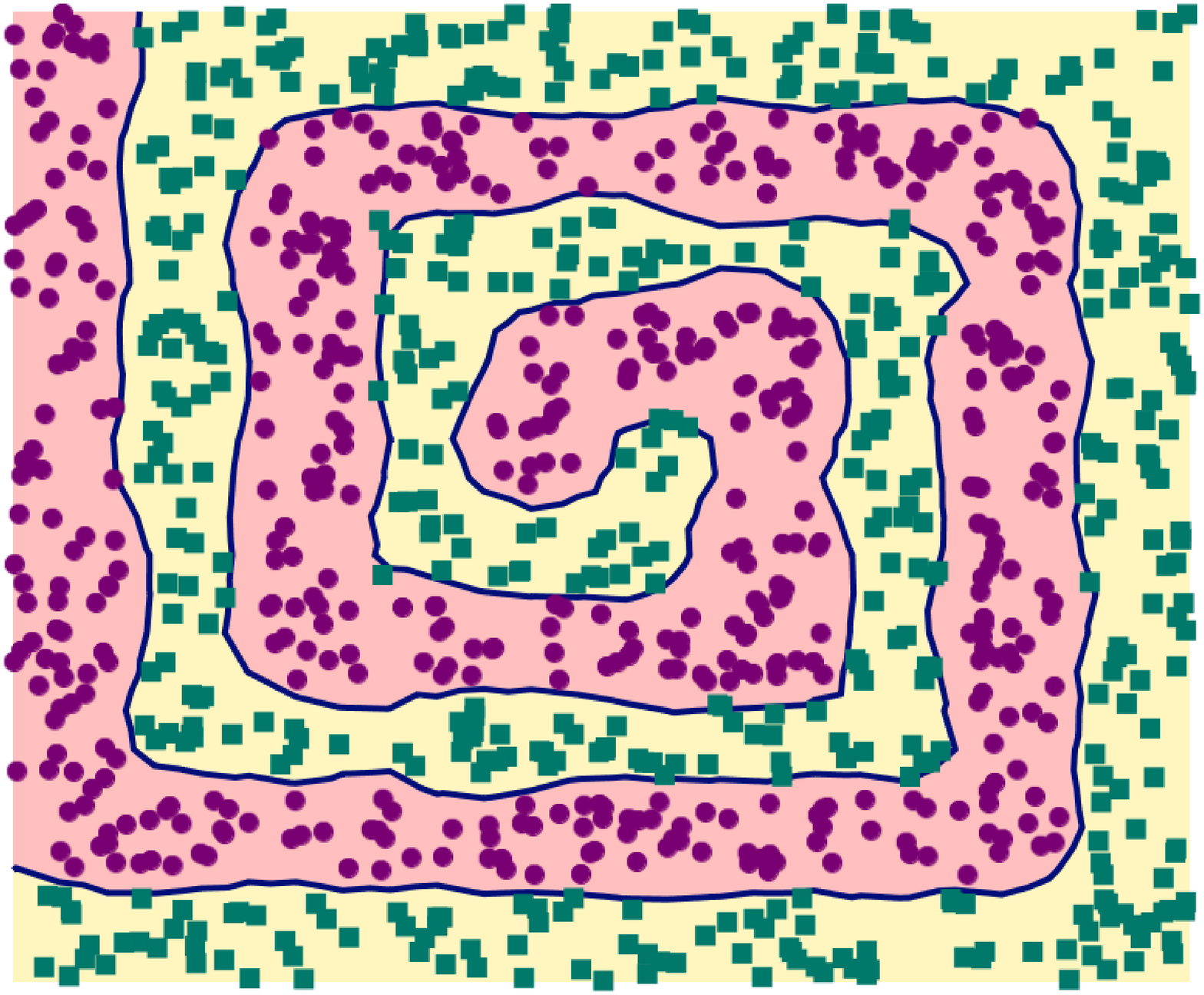}
}\hspace{-.21cm}
\subfigure[]{
\includegraphics[height=1.7cm,width=2cm]{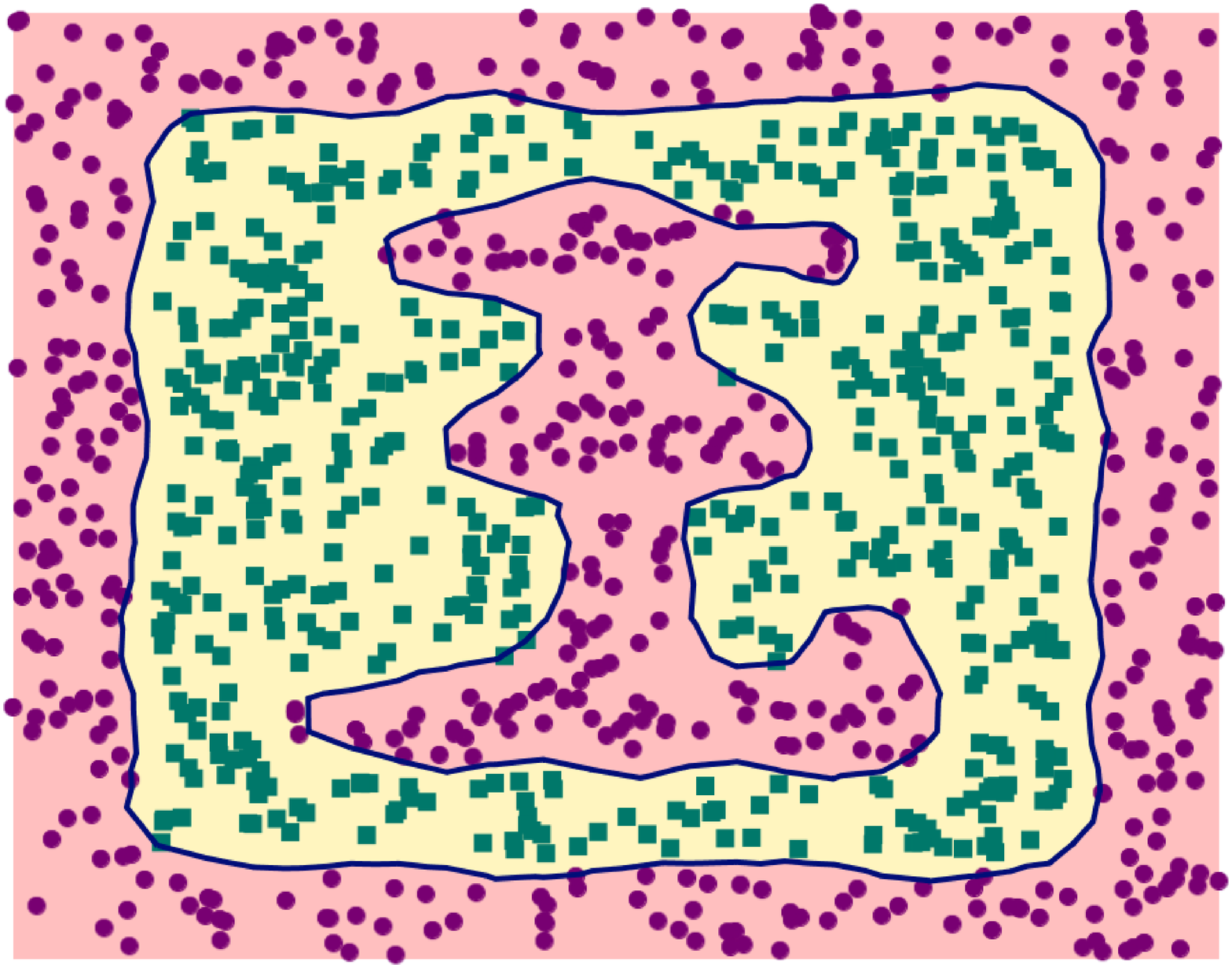}
}
\caption{Use quantum RBF network to solve binary classification problems:
1024 training samples, 10 weight parameters are used.}
\label{test1}
\end{figure}

We tested dozens of examples with different types of classification patterns.
In Figure \ref{test1}, we list four typical testing results.
They indicate that the quantum RBF network can solve the binary classification problems with high precision.

\begin{figure}[ht]
     \centering
     \subfigure[RCP vs. $m$]{
     \includegraphics[height=3cm,width=4cm]{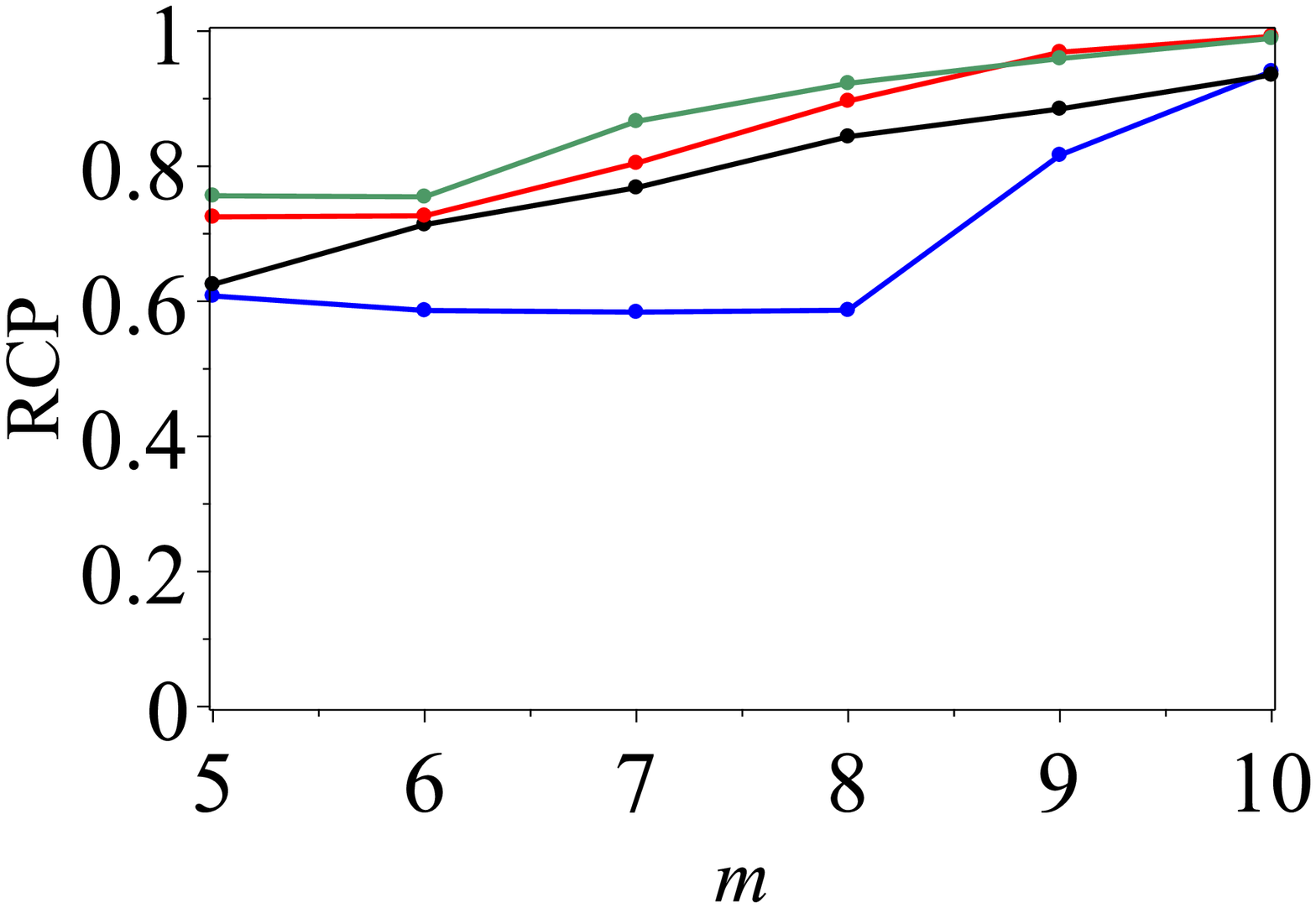}
     \label{test3-a}
     }
     \subfigure[MSE vs. $m$]{
     \includegraphics[height=3cm,width=4cm]{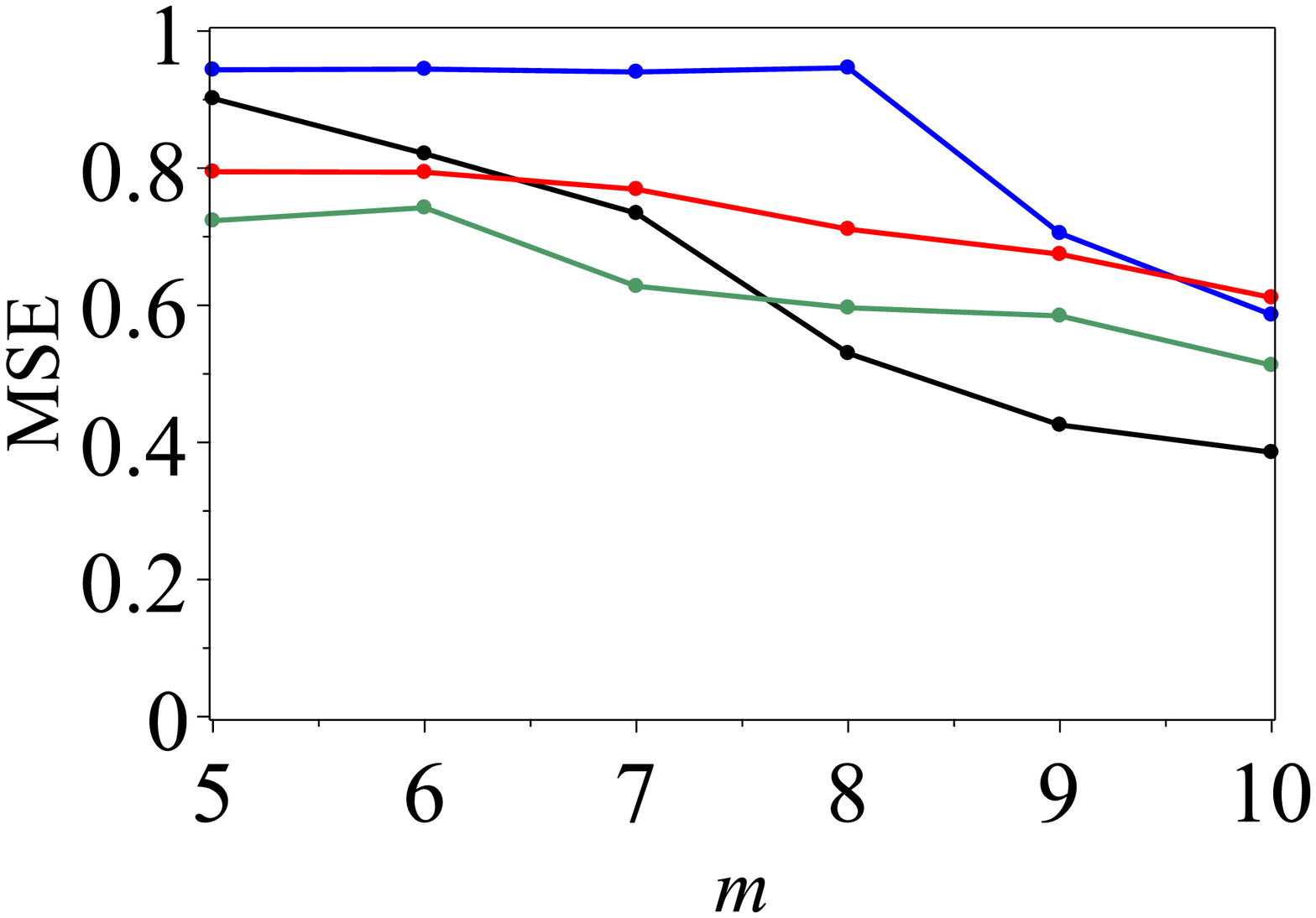}
     \label{test3-b}
     }
    \caption{The influence of the number of training samples on the performance of the quantum RBF network: $m$ is the logarithm of the number of training samples, the blue, black, red and green curves respectively represents the results of pattern (a), (b), (c), (d) in Figure \ref{test1}.}
    \label{test3}
\end{figure}

Usually, the performance of a machine learning method
improves with the increment of training samples.
In Figure \ref{test3}, we consider the influence of the number of training samples on the performance of the quantum RBF network.
Note that for classifiers, during training we concern
more about the ratio of correct predictions (RCP), which is defined by
\be
{\rm RCP} := 1 - \frac{1}{4M} \sum_{t=1}^M \Big( {\rm sign} (w(\vec{\theta}) \cdot f(x^{(t)})) - r^{(t)}\Big)^2.
\ee
It computes the ratio of training samples that  being given the correct labels by the neural network after training.
For a classifier, we hope its RCP is large. However, the training is accomplished by minimizing the mean square error (MSE) defined by equation (\ref{error1}).
If MSE is small, then
this classifier can also be used to approximate functions, that is it can solve regression problems as well.

In Figure \ref{test3} we tested all the four patterns of Figure \ref{test1}. In each situation, we randomly test 10 examples,
then compute the average values of the RCPs and MSEs.
As we can see from Figure \ref{test3-a}, the overall trend of RCP is
increasing. When using 1024 training samples, the RCPs are all larger than 0.9. This means that the performance of the quantum RBF network
to solve classification problems increases when the number of training samples increases.
On the other hand, the overall trend of MSE shown in Figure \ref{test3-b} is decreasing.
However, the MSEs are still large ($\geq 0.35$) even if we use 1024 training samples.
As a result, this kind of neural network seems not a good method to solve regression problems.

Finally, in Table \ref{table} we compare the performance of
the quantum RBF network and the classical
RBF network defined in Section \ref{Radial basis function network} in solving 10 different classification problems.
All the tests are performed on Maple 2019 with command \texttt{Minimize} in the package \texttt{Optimization}. The processor of the DELL computer is Intel(R) Core(TM) i3-6100 CPU@3.70GHz.

{\renewcommand
\arraystretch{1.2}
\begin{table}[h]\centering
\begin{tabular}{|c|c|c|c|c|c|} \hlinewd{1pt}
~NO~ & ~~~~ & ~Time(sec.)~  & RCP   & MSE  & INF \\ \hlinewd{1pt}
\multirow{2}*{1}  & Q & 47.32 & 0.998  & 0.465   & 0.998 \\ \cline{2-6}
    & C & 72.89 & 0.992 & $2.468\times 10^{-11}$ & 0.996 \\ \hlinewd{.7pt}
\multirow{2}*{2}   & Q & 37.78 & 0.998  & 0.742  & 0.992 \\ \cline{2-6}
    & C & 70.58 & ~0.992~ & ~$3.906\times 10^{-15}$~ & ~0.988~ \\ \hlinewd{.7pt}
\multirow{2}*{3}  & Q & 56.57 & 0.953  & 0.291  & 0.963  \\ \cline{2-6}
    & C & 118.42 & 0.998 & $4.715\times 10^{-3}$ & 0.955  \\ \hlinewd{.7pt}
\multirow{2}*{4}  & Q & 319.42 & 0.959 & 0.567 & 0.951   \\ \cline{2-6}
   & C & 3151.39 & 0.998 & $2.717\times 10^{-7}$ & 0.947   \\ \hlinewd{.7pt}
\multirow{2}*{5}  & Q & 394.91 & 0.949 & 0.231 & 0.924   \\ \cline{2-6}
   & C & 3284.71 & 0.996 & $4.912\times 10^{-3}$ & 0.959   \\ \hlinewd{.7pt}
\multirow{2}*{6} & Q & 456.61 & 0.971  & 0.686 & 0.879 \\ \cline{2-6}
   & C & 3091.18 & 0.998 & $2.089\times 10^{-7}$ & 0.887 \\ \hlinewd{.7pt}
\multirow{2}*{7}  & Q & 8025.65 & 0.941  & 0.586 & 0.968 \\ \cline{2-6}
   & C & / & / & / & / \\ \hlinewd{.7pt}
\multirow{2}*{8}  & Q & 7951.36 & 0.935  & 0.385 & 0.989 \\ \cline{2-6}
   & C & / & / & / & / \\ \hlinewd{.7pt}
\multirow{2}*{9} & Q & 7365.29 & 0.992  & 0.611 & 0.992 \\ \cline{2-6}
   & C & / & / & / & / \\ \hlinewd{.7pt}
\multirow{2}*{10} & Q & 7133.19 & 0.989  & 0.513 & 0.947 \\ \cline{2-6}
   & C & / & / & / & / \\ \hlinewd{1pt}
\end{tabular}
\caption{
Comparison of the performance of the quantum and classical RBF networks. In the second column, Q and C respectively refer to quantum RBF network and
the classical  RBF network.}
\label{table}
\end{table}
}

In Table \ref{table}, INF refers to the ability
to do inference on new data after training.
It has a similar definition to RCP.
In each case, we use $M$ new samples to compute this quantity.
The first three have simple patterns, so we just use $M=256$ training samples. The pattern of the next three
are a little complicated, thus $M=512$  training samples are used.
The last four are the four patterns shown in
Figure \ref{test1}, where 1024 training samples are used. In the last four examples, the
backslash ``$/$" means that the classical RBF network cannot finish the
training after 10 hours.
From the time used in each case, we can train the quantum RBF network much faster than the classical RBF network.
The difference becomes more clear
when more training samples are used. This is consistent with Theorem \ref{thm}.
When both methods work,
the RCP and INF have little difference. This shows that
the performance of these two methods in solving classification problems are close to each other.
However, the MSEs of the quantum
RBF network are always large.
So it is not a good method to solve regression
problems.

\section{A related idea towards the support vector machines}
\label{Some related ideas toward the support vector machines}

Support vector machine (SVM) is an important kernel
learning method \cite{cortes1995support}.
In this section, we present a generalization of
the idea used in RBF network to SVM.
This can improve the efficiency of training SVM.
But it removes the convex optimization theory of SVM
and cannot find the support vectors.
So this method is preferable when the computation
is an obstacle to SVM.

Consider the binary classification problem
discussed in equation (\ref{samples}).
Let $f$ be a feature map, then the kernel-based SVM
is a method aims to find a hyperplane
$f(x)\cdot v-b=0$ that can separate the
two classes with maximal margin. Points on
the boundaries  $(f(x^{(t)})\cdot v-b)r^{(t)}=1$
are called support vectors. The distance
from support vectors to the hyperplane
$f(x)\cdot v-b=0$ equals $1/\|v\|$.
Thus SVM reduces to solve the following optimization problem:
\be\ba{rll} \label{svm-1} \vspace{.2cm}
{\rm Min} & \ds \frac{1}{2}\|v\|^2 \\
{\rm s.t.} &
(f(x^{(t)})\cdot v-b)r^{(t)} \geq 1,~
t=1,2,\ldots,M.
\ea\ee

Introduce the Lagrangian function
\be \label{svm-2}
L = \frac{1}{2}\|v\|^2 - \sum_{t=1}^M w_t \Big((f(x^{(t)})\cdot v-b)r^{(t)}-1\Big).
\ee
Setting the partial derivatives $\partial L/\partial v$ and $\partial L/\partial b$ as zero leads to
\be  \label{svm-3}
v =  \sum_{t=1}^M w_t r^{(t)}  f(x^{(t)}),
\quad
\sum_{t=1}^M w_t r^{(t)} = 0.
\ee
Substituting them into equation (\ref{svm-2}), then
we obtain the Lagrangian dual problem of (\ref{svm-1}) as follows
\be \ba{rll} \label{svm-4}
{\rm Max} & \ds Q(w) = \sum_{t=1}^M w_t - \frac{1}{2} \sum_{s,t=1}^M w_s w_t r^{(s)} r^{(t)}  f(x^{(s)}) \cdot f(x^{(t)}), \\ \vspace{.2cm}
{\rm s.t.} & \ds \sum_{t=1}^M w_t r^{(t)} = 0, \quad w_t \geq 0,~
t=1,2,\ldots,M.
\ea\ee

The solution of the problem (\ref{svm-4}) is usually sparse, that is $w_t\neq 0$ if $x^{(t)}$ is a support vector. By equation (\ref{svm-3}), $v$ is a linear combination of the support vectors. The decision boundary is
\be \label{decision boundary}
\sum_{x^{(t)}:~\rm support~vectors} w_t r^{(t)}  f(x^{(t)}) \cdot f(x) - b = 0.
\ee

The SVM can be described in the language of RBF network.
The support vectors are the centers used in the hidden layer, and $b$ is the partial bias. Since we do not know the support vectors in advance, if we simply use all the training samples in equation (\ref{decision boundary}), then the decision boundary is the same as that defined in the RBF network.
However, the approaches to find the weights are different in these two methods.

\begin{figure}[h!]
     \centering
     \subfigure{
     \includegraphics[height=3cm,width=4cm]{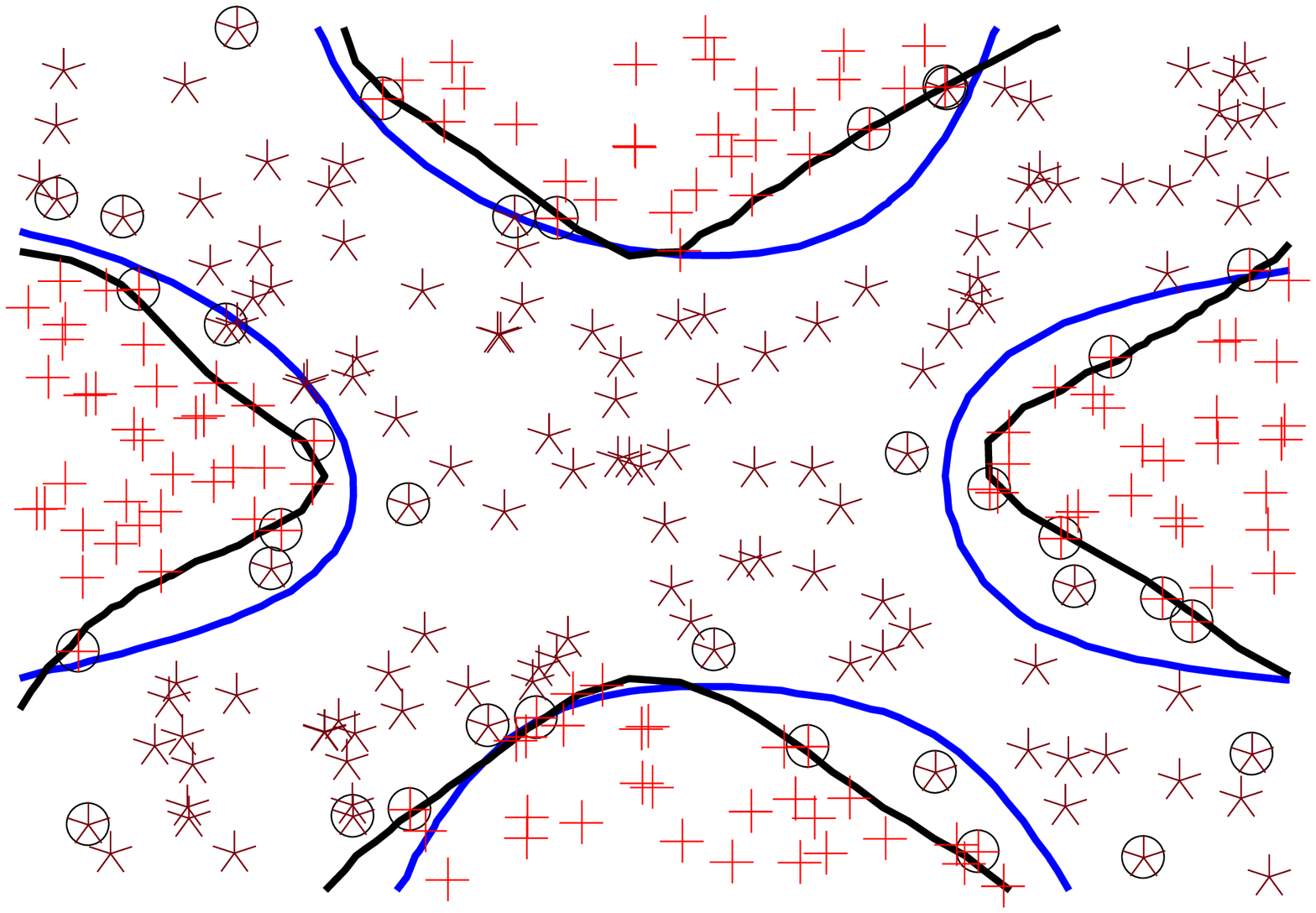}
     \label{test4-a}
     }
     \subfigure{
     \includegraphics[height=3cm,width=4cm]{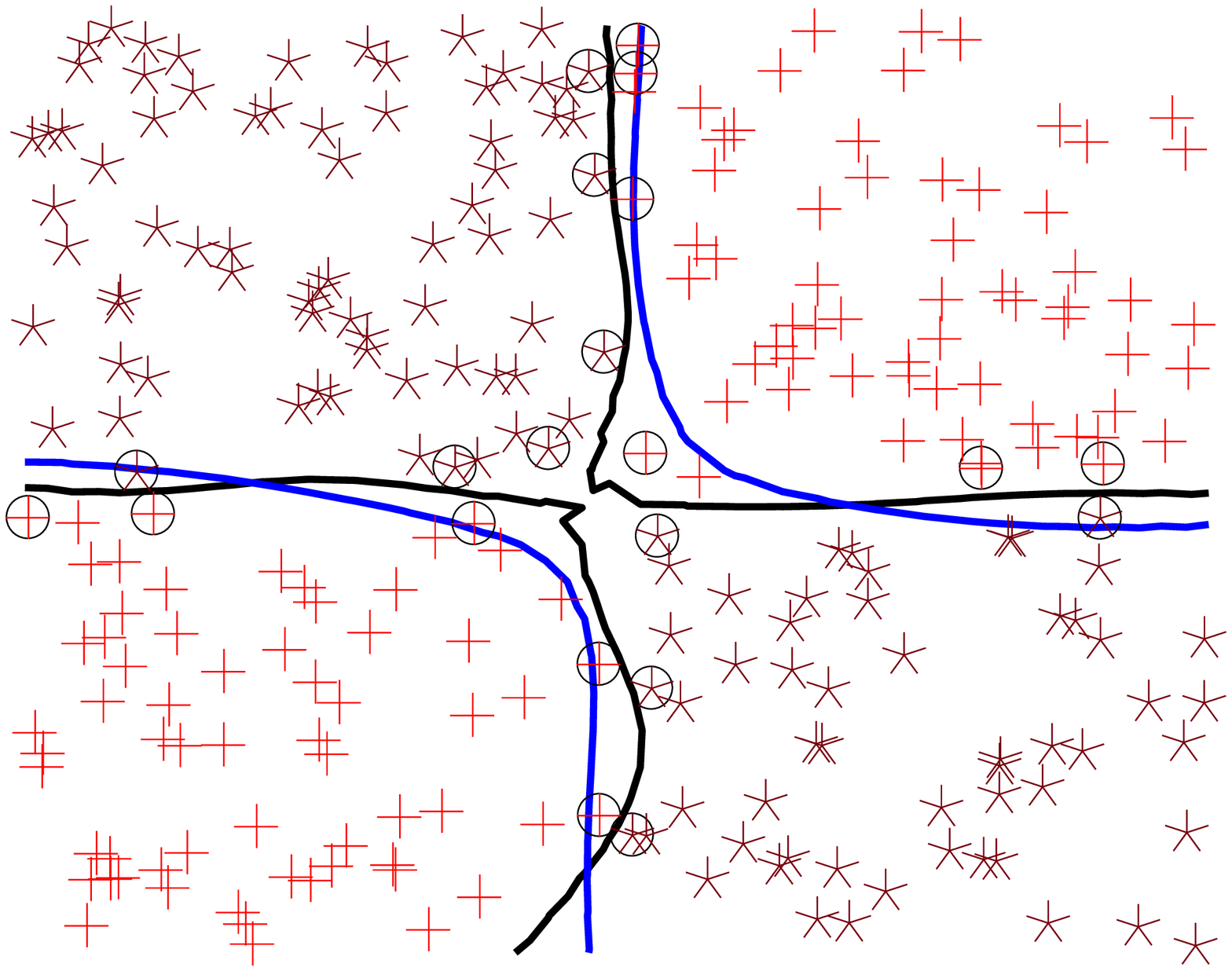}
     \label{test4-b}
     }
    \caption{The black curves are the decision boundaries discovered by the SVM with the usual choice of the weight;
    the blue ones are obtained by choosing the weight has the form (\ref{weight1}). The circles represent the support vectors.}
    \label{test4}
\end{figure}

The advantage of SVM is the convex optimization theory it relies on. This perhaps makes it the most elegant of all kernel-based learning methods.
The major limitation of SVM is the fast increase in
the computing and storage requirements with respect to the number of training examples.
Based on the idea of quantum RBF network, we can change the weight $w=(w_1,\ldots,w_M)$ into the tensor product form (\ref{weight1}).
The advantage of doing so is that
only $\log M$ parameters are required. This can save a lot of  time to solve the optimization problem.
However, it will change the nature of SVM. Moreover,
the convex property disappears. Numerical tests show that this method can solve classification problems with high
precision (see figure \ref{test4} for two examples), while it cannot find the support vectors.
The solution found by this method is not sparse. This kind of idea is preferable when
the training of the SVM is inefficient.
When we have a large-scale quantum computer,
the efficiency of training can be further improved.



\section*{Acknowledgement}

This work was supported by the QuantERA ERA-NET Cofund in Quantum Technologies implemented within the European
Union's Horizon 2020 Programme (QuantAlgo project), and EPSRC grants EP/L021005/1 and EP/R043957/1.

\nocite{*}
\bibliography{apssamp}

\end{document}